\documentstyle[12pt,aasms4]{article} 
\begin{document} 
 
\title{The  Direct Test of Cosmological Model for Cosmic Gamma-Ray Bursts  
Based on the Peak Alignment Averaging} 
 
\author{Igor G.~Mitrofanov, Maxim L.~Litvak and Dmitrij A. Ushakov}  
 
\affil{Space Research Institute, Profsojuznaya str. 84/32, 117810 
Moscow, Russia} 
 
\begin{abstract} 
 
The cosmological origin of cosmic gamma-ray bursts is tested using the method  
of peak alignment for the averaging of time profiles. The test is applied to the  
basic cosmological model with standard sources, which postulates that  difference  
between bright and dim bursts results from different cosmological red-shifts of   
their sources. The average emissivity curve (ACE$_{bright}$) of the group of  
bright BATSE bursts is approximated by a simple analytical function, which takes  
into account the effect of squeezing of the time pulses with increasing energy of  
photons. This function is used to build the model light curve for ACE$_{dim}$ of  
dim BATSE bursts, which takes into account both the cosmological time- 
stretching of bursts light curves and the red-shifting of photons energies. Direct  
comparison between the  model light curve and the ACE$_{dim}$ of dim bursts  
is performed, which is based on the estimated probabilities of differences between  
ACEs of randomly selected groups of bursts.  It shows no evidence for the  
predicted cosmological effects. The $3 \sigma$ upper limit of the  average red- 
shift $z_{dim}$ of emitters of dim bursts is estimated to be as small, as $\sim  
0.1-0.5$, which is not consistent with values $\sim 1$ predicted by the  known  
cosmological models of gamma-ray bursts.

{\it Subject headings}: cosmology: theory-gamma rays: bursts
 
\end{abstract} 
 
\section{Introduction} 
 
The cosmological model of cosmic gamma-ray bursts is commonly accepted, as  
one of the most promising concept of the origin of gamma-ray bursts (GRBs).  
However, it has not been finally approved yet by the observational data. Two  
critical tests were suggested to verify the basic model with standard cosmological  
sources: dim bursts have to be time-stretched and red-shifted in comparison with   
bright events.  
 
Gamma-ray bursts are known to have very different time histories, and one  
hardly could check the cosmological effects by direct comparison between  
particular events. These tests should be based on some averaged time-based and  
spectrum-based signatures, which represent the basic properties of GRBs.  
Several statistical tests have already been implemented to compare different  
groups of GRBs and to resolve the predicted cosmological effects. In the case of  
time-dilation, two scientific groups have checked the average emissivity  
curves (ACEs) derived from the peak alignment averaging of bright and dim  
bursts, and came to opposite conclusions: time-dilation of dim bursts was seen  
by one group (Norris {\it et al.} 1994a, 1994b, 1996; Bonnel {\it et al.} 1996) and it was not seen by 
another  
one (Mitrofanov {\it et al.}, 1992a,b, 1994, 1996). Possible reasons for  
disagreement were discussed (Band 1994, Mitrofanov {\it et al.} 1996), and the  
tentative conclusion has been drawn that the claimed dilation of dim bursts was  
possibly resulted from some systematic in separation of bright and dim groups  
of events.  
 
On the other hand, in the case of spectral red-shift, all groups involved have a  
consensus that bright GRBs have larger averaged spectral hardness than dim  
events. It was named, as the effect of hardness/intensity correlation  
(Mitrofanov {\it et al.} 1992b,c, 1994, 1996; Paciesas {\it et al.} 1993, Norris {\it  
et al.} 1994a,b). While this effect was originally seen for the average hardness  
ratio (defined as the ratio of counts at high and low energy channels), recently it  
was found also for the average peak energy $<E_{p}>$ of ${\nu F_{\nu}}$  
energy spectra (Mallozzi {\it et al.} 1995). The average peak energies  
$<E_{p}>$ of the integral spectra of photons were found to correlate with the  
photons¯ peak fluxes $F_{max}^{(256)}$ at the 256 ms time scale. The effect of  
hardness/intensity correlation  may be interpreted, as a result of cosmological red- 
shift of dim gamma-ray bursts in respect to bright events. The corresponding  
cosmological red-shift factor about 1.6-2.2 (Mallozzi {\it et al.} 1995) is  
consistent with original cosmological models based on the interpretations of log  
N - log F distribution (e.g. see Emslie and Horrack 1994). 
 
Therefore, there is a discrepancy between different groups about time-dilation of  
dim GRBs in respect to bright bursts, but, on the other hand,  there is an  
commonly accepted agreement between them for the hardness/intensity  
correlation. 
 
Separate pulses of GRBs are known to squeeze with increasing energy of  
photons (Norris {\it et al.} 1986, Fenimore {\it et al.} 1995), and, therefore, the  
average curve of  
emissivity becomes narrower at higher energies  (Mitrofanov {\it et al.} 1996).  
According to the cosmological model, when bright and dim bursts are detected at  
the same energy band in the observer frame of reference, their time profiles were  
actually emitted at less hard and more hard energy ranges at comoving frames of  
reference, respectively. Therefore, making a comparison of bright and dim bursts  
from sources with small and large red-shifts, one should suppose that the intrinsic  
squeezing of the light curves of dim bursts due to the increase of energy of the  
emitted  
photons could partially compensate their stretching due to the cosmological time- 
dilation.  
 
This paper provides a test of the basic cosmological models of GRBs assuming  
them to be standard sources. It uses the average emissivity curves for groups of  
bright and dim bursts and takes into account the effects of cosmological time- 
stretching in the observer frame together with the internal energy dependent  
squeezing of bursts light curves in the comoving frames. 
 
\section{Analytic Approximation of the Average Curve of Emissivity for Group  
of Bright GRBs} 
 
The average curve of emissivity (ACE) of GRBs was introduced (Mitrofanov  
{\it et al.} 1994, 1996), as a general signature of bursts time variability. To build  
an ACE, all time histories of averaging bursts should be normalized by peak  
numbers of counts $C_{max}$, then should be aligned at their peak bins  
$t_{max}$ and then should be averaged at each of another bins. Comparison  
between the First, the Second and the Third BATSE Catalogs (Fishman {\it et  
al.} 1994, Meegan {\it et al.} 1994 and Meegan {\it et al.} 1995) has shown that  
ACE has rather stable shape: it has one peak profile with steep rise front and  
gentle back slope, and its width decreases with increasing energy of photons used for averaging
(Mitrofanov {\it et al.} 1994, 1996).  
 
For the present analysis the DISCLA data were used from the large area BATSE  
detectors (LADs) with 1024 ms time resolution on three discriminator channels,  
number 1 (25-50 keV), number 2 (50-100 keV) and number 3 (100-300 keV).  
Two basic intensity groups of BATSE GRBs were selected from the Third BATSE  
Catalog (3B) (Meegan {\it et al.} 1995): 296 bright bursts with  
$F_{max}^{(1024)}>1$ photons cm$^{-2}$ s$^{-1}$ and  332 dim events with  
$F_{max}^{(1024)}<1$ photons cm$^{-2}$ s$^{-1}$. Only bursts with  
$t_{90}>1.0$s were taken into account for consideration.  
 
The group of bright bursts is used as the reference sample to find the analytical  
approximation of ACE$_{bright}$ at different discriminator channels (Figure 1).  
The function 
 
\begin{equation}    
f_{bright}^{(i)}(t)=(    
\frac{t_{0}^{(i)}}{t_{0}^{(i)}+|t -t_{max}|}    
)^{{a_{RF}^{(i)}}, {a_{BS}^{(i)}}}      
\label{ep1}   
\end{equation}    
  
\noindent approximates ACE$_{bright}^{(i)}$ profiles at each discriminator  
channel $i=1,2,3$ with different power  
indexes $a_{RF}^{(i)}$ at the rise front (RF) $t<t_{max}$  and $a_{BS}^{(i)}$  
at the back slope (BS) $t>t_{max}$, respectively. Instead of three different  
functions (1) for each of three channels, a single function $f_{bright}(t, E)$ could  
be implemented, which approximates the shape of ACE$_{bright}$  at different  
energies $E$, which correspond to these channels 
 
\begin{equation}    
f_{bright}(t,E)=(    
\frac{    
t_{bright}(E)    
}    
{    
t_{bright}(E)+ | t -t_{max} |    
}    
)^{{a_{RF}(E)},     
{a_{BS}(E)}},      
\label{ep2} 
\end{equation}    
 
\noindent where the functions  
 
\begin{equation}    
t_{bright}(E)=t_{bright}^{(0)} \cdot (E/173~keV)^{\alpha_{1}}    
\label{ep4} 
\end{equation}    
\begin{equation}    
a_{RF}(E)=a_{RF}^{(0)} \cdot (E/173~keV)^{\alpha_{2}}    
\label{ep5} 
\end{equation}    
\begin{equation}    
a_{BS}(E)=a_{BS}^{(0)} \cdot (E/173~keV)^{\alpha_{3}}    
\label{ep6} 
\end{equation}    
 
\noindent represent the change of ACE$_{bright}$ shape with energy. A  
difference between three observed ACE$_{bright}^{i}$ profiles (Figure 1) and  
the model approximation (2) could be evaluated using the function 
 
\begin{equation}    
S_{bright}= 
\displaystyle 
\sum_{i} 
\sum_{j}     
\frac{    
 (ACE_{bright}^{(i,j)}- f_{bright}(t_{j},     
E_{i}))^{2}    
}    
{    
\sigma^{2}(ACE_{bright}^{(i,j)})    
}    
\label{ep7} 
\end{equation}

\noindent where $E_{i}$ corresponds to mean energies at three discriminator  
channels  
$i=1,2,3$ and $t_{j}$ corresponds to the time bins of ACE curves from $j= -19$  
up  
to $j=+19$. Errors of observed ACE profiles were estimated from the sample  
variance. 
 
The parameters of approximation $t_{bright}^{(0)}= 1.80\pm^{0.33}_{0.28}$ s,  
$a_{RF}^{(0)}=1.31\pm^{0.13}_{0.12}$,  
$a_{BS}^{(0)}=1.10\pm^{0.10}_{0.09}$, $\alpha_{1}= -0.10 \pm 0.16$,  
$\alpha_{2} = 0.06 \pm 0.09$ and  $\alpha_{3}=0.11 \pm 0.08$ were estimated  
from the best fitting of all three ACE$_{bright}^{(i)}$ profiles at channels  
$i=1,2,3$. This fitting leads to the minimum $S_{bright}^{(min)}$ of  Exp.(6),  
which corresponds to rather small value of the Pearson criterion: reduced  
$\chi^{2}=0.66$ for 108 degrees of freedom. Therefore, one might conclude that  
Exp. (2) gives a rather good approximation of the observed ACE$_{bright}$  
profiles for the basic group of bright bursts at a broad energy range from 25 up to 300  
keV. On the other hand, rather small values of the reduced $\chi^{2}$ points out  
that the errors of  
ACE$_{bright}$ was probably overestimated by the sample variance algorithm,  
or there were some  correlation between them. 
 
However, the Pearson criterion allows to determine the confidence region for the  
estimated  
parameters of the fitting function (2). According to Lampton {\it et al.} (1976),  
the confidence region for the significance level ${\lambda}$ could be determined  
by the 5-dimensional contour $S_{countur}$ in the 6-dimensional parameter  
space, which is given by the equation 
 
\begin{equation} 
S_{countur} = S_{bright}^{(min)} + \chi^{2}_{6}(\lambda), 
\end{equation} 
 
\noindent where  $\chi^{2}_{6}(\lambda)$ represents the value of $\chi^{2}$  
distribution  
for significance ${\lambda}$ for 6 degrees of freedom. Errors ${\pm 1\sigma}$  
for each of six parameters, as presented above, were estimated from the  
condition that Exp. (6) for $S_{bright}$ becomes equal to $S_{countur}$  
when the parameter goes up and down from the best fitting value, while another  
five parameters are used as free parameters for minimization.  Therefore, each of  
these 12 points could be interpreted as ${\pm\sigma}$ deviations from the  
minimum point along the axes of corresponding parameter inside a 5-dimensional  
contour $S_{countur}$. 
 
Actually, these 12 points in the six-dimensional parameter space correspond to   
12 fitting models (2) of the ACE$_{bright}^{i}$ curves. Were taken all together,  
they would present the ${\pm 1\sigma}$ ${\it  
corridor}$ of analytical approximations around the best fitting model, which leads  
to $S_{bright}=S_{bright}^{(min)}$.  
The boundary curves of this corridor are presented at Figure 1. One might see  
that all these models provide rather good approximation of all three  
ACE$_{bright}^{(i)}$ profiles measured at three energy discriminitors.  
 
\section{Comparison between the average emissivity curves for different groups  
of bursts} 
 
Particular gamma-ray bursts are known to have very different time histories and  
energy spectra. Therefore, ACE curves could be different for particular groups  
of bursts randomly selected from the total data base. Groups with $N_{rep}$  
bursts could be defined, as representative samples, provided the differences  
between  
their ACEs would be comparable with the errors from the sample variance for  
each group. For smaller groups with $N<N_{rep}$ a difference  
between ACE curves could be significantly larger than it would be expected from  
the sample variance. Therefore, the comparison of ACE of different groups has to  
take into account the actual distribution of differences between  
ACEs profiles due to a random choice of contributing bursts. 
 
Nobody knows how large is the representative sample of time histories of GRBs,  
but it seems from the comparison of ACE curves for 1B, 2B and 3B  
databases that $N_{rep}$ could be about the presently available number of bursts  
$\sim10^{3}$ (Mitrofanov {\it et al.} 1997). As it was found there, the Pearson  
criterion provides a rather  sensitive test to measure a difference between  ACEs  
for any two groups of bursts, namely groups I and II, at any discriminator channel   
$i$:  
 
\begin{equation} 
S_{(I-II)}^{(i)}= 
\sum_{j} 
\frac{ (ACE_{(I)}^{(i,j)}- 
ACE_{(II)}^{(i,j)})^{2}} 
{\sigma^{2}(ACE_{(I)}^{(i,j)})+\sigma^{2}(ACE_{(II)}^{(i,j)}) 
} 
\end{equation} 
 
The magnitude $S_{(I-II)}$ was used to compare ACEs profiles for randomly  
selected groups of events. It was found that groups with  
$N$ increasing from $\sim 30$ up to $\sim 300$ become more and  
more representative with respect to the full set. In particular, the probability  
distribution $P_{300}$ of  $S_{(I-II)}^{(2)}$ at discriminator $i=2$ was  
obtained from $10^{5}$ random choices of two groups with N=303 among the total 3B set of 638 
BATSE bursts (Figure 2). This  
distribution does not depend significantly on the intensity of selected bursts,  
because the main contribution into $S_{(I-II)}$ comes from the actual difference  
of their time histories. 
 
Thus, Exp. (8) could be used for direct comparison between ACE profiles  
of groups of bright and dim bursts, and the significance of a physical difference  
$S$ between  
them could be estimated as the probability of obtaining $S$ greater than $S_{(I- 
II)}$ according to the distribution $P_{300}(S_{(I-II)})$  
provided by the Monte Carlo random choice test (Figure 2). This probability  
distribution will be used below to compare the analytical  
model based on the ACE$_{bright}$ of bright bursts and the actual  
ACE$_{dim}$ measured for the group of dim events. 
 
\section{Direct cosmological Test Based on the Analytic Approximation of the  
Average Curve of Emissivity} 
 
The simplest test of cosmological model of GRBs could be based on the {\it  
standard candle} assumption, which means that everywhere at cosmological  
distances all sources have the same properties in their comoving frame.  
This basic version of the cosmological model assumes that all groups of bursts,  
provided would be averaged in comoving frames, should have the same ACEs.  
Therefore, any difference between ACEs of bright and dim bursts measured in  
the observer frame should point out on the cosmological effects.  
 
Let us assume that the emitters of bright and dim bursts  
have average red shifts $z_{bright}$ and $z_{dim}$, respectively. If two  
standard sources at $z_{bright}$ and $z_{dim}$ emit bright and dim bursts with  
photons energy $E_{0}$ and variability time scale $\tau_{0}$, they would be  
detected in the observer frame at energies $E_{bright}=E_{0}/(1+z_{bright})$  
and $E_{dim}=E_{0}/(1+z_{dim})$ and with variability at time scales  
$\tau_{bright}=\tau_{0}(1+z_{bright})$ and  
$\tau_{dim}=\tau_{0}(1+z_{dim})$,  
respectively. The so-called stretching factor could be introduced 
 
\begin{equation} 
Y(z_{bright}, z_{dim}) = (1 + z_{dim})/(1 + z_{bright}),   
\end{equation} 
 
\noindent which equals to the ratio of energies of photons $E_{bright}/E_{dim}$  
and/or to  
the ratio of variability time scales $\tau_{dim}/\tau_{bright}$ at the observer  
frame of  
reference, provided they were the same in comoving frames.  
 
To test the basic cosmological model, the analytical approximation   
$f_{bright}(t,E)$ (Exp. (2)) should be transformed into the model function  
$f_{dim}(t,E)$ according to cosmological red-shifting and time-stretching  
transformations, which  
has to represent the measured ACE$_{dim}$ profiles for the group of dim bursts.  
According to the assumption of standard candles one should postulate 
 
\begin{equation} 
f_{dim}(t,E)=f_{bright}(\frac{t}{Y}, E\cdot Y).      
\end{equation} 
  
Using the Exp. (2), one might represent Exp. (9), as the following 
 
\begin{equation} 
f_{dim}(t,E; Y)=(\frac{Y\cdot t_{bright}(Y\cdot E)} 
{Y\cdot t_{bright}(Y\cdot E)+ | t  
-t_{max} |})^{a_{RF}(Y\cdot E), a_{BS}(Y\cdot E)},  
\end{equation} 
 
\noindent which could be used either as a function of one stretching parameter  
$Y$, or as a function of two red-shifts $z_{bright}$ and $z_{dim}$. 
 
Figure 3 presents ACE$_{dim}^{(i)}$  profiles for  the basic group of 332 dim bursts  
from the 3B database with  $F_{max}^{(1024)}<1$ photons cm$^{-2}$ s$^{- 
1}$ observed at  
three energy discriminators with numbers $i=1,2,3$. Expression (11) provides a  
trial function for the ACE$_{dim}^{(i)}$ profiles with the factor $Y$, as a free  
parameter. To compare the model with observations, the function  
$S_{dim}$ could be used similar to $S_{bright}$ (6). Table 1 presents the best  
fitting values $Y^{*}$ for each of the three ACE$_{dim}^{(i)}$ profiles fitted  
separately, and one more value for the joint fit of all three curves together. The  
errors of $Y^{*}$ correspond to the range of the best fitting values of $Y$ for  
the 12 different  
models (11) based on the initial model (2) with ${\pm\sigma}$ deviations of its  
six parameters (see Section 2). 
 
The values of minima $S_{dim}^{(min)}$ for the best fitting parameters  
$Y^{*}$ are rather large, and according to the Pearson criterion, the model of  
equation (11)  
does not agree with the observed ACE$_{dim}^{(i)}$ profiles for discriminators  
$i=1, 3$ and $(1+2+3)$. Only in the case of discriminator $i=2$ the model (11)  
with $Y^{*}$=0.85-0.87 formally agrees with the ACE$_{dim}^{(2)}$ profile.  
Moreover, instead of the expected $\it stretching$, all the best fitting factors  
$Y^{*}$ (Table 1) correspond  to ${\it squeezing}$ of ACE$_{dim}^{(i)}$  
profiles with respect to the analytic approximation of  ACE$_{bright}^{(i)}$ for  
bright bursts (Exp. (2)).  
 
However, the classical Pearson criterion based on the $\chi^{2}$-distribution  
could not be applied in this case, because it does not take into account the actual  
distribution of differences $S_{(I-II)}$ between ACEs profiles due to a random  
selection of contributing events. A more accurate test of the basic cosmological  
model is done below, which takes into account the probability distribution of   
$S_{(I-II)}$  resulting from the random sampling of BATSE bursts (see  
Section 3). This test has to provide the upper limits of $z_{dim}$ for the  
basic cosmological model with standard sources, which could be deduced from  
the observed profiles of ACE$_{bright}$ and ACE$_{dim}$. 
 
According to this model, a group of bursts with fluxes $\sim F$ corresponds to a  
definite red-shift $\sim z$. While in the Euclidean space there is a flux dilution  
law $\sim R^{-2}$, which establishes the well-known flux/distance relation for  
standard sources, the non-Euclidean dilution of fluxes from cosmological emitters  
is influenced by the effects of photon energy red-shifting and light curve time- 
stretching.  
 
While each burst has a particular energy spectrum, the average spectral  
distribution could be obtained for any selected group of bursts as well as ACEs  
were obtained for their time histories.  According to Band {\it et al.} 
(1993), the energy spectra of  BATSE bursts $\phi(E)$ could be described by the  
law 
 
\begin{equation} 
\phi(E)=A\cdot (\frac{E}{100~keV})^{\alpha}\cdot e^{- 
\frac{E(2+\alpha)}{E_{peak}}}      
\end{equation} 
\noindent if 
\begin{equation} 
E<(\alpha-\beta)\cdot\frac{E_{peak}}{(2+ \alpha)}   
\end{equation} 
\begin{equation} 
\phi(E)=A\cdot ((\alpha- 
\beta)\cdot\frac{E_{peak}}{100~keV(2+\alpha)})^{(\alpha-\beta)}\cdot  
e^{\frac{E_{peak}}{100~keV}}\cdot (\beta-\alpha)^{\beta}      
\end{equation} 
\noindent if 
\begin{equation} 
E>(\alpha-\beta)\cdot\frac{ E_{peak}}{(2+ \alpha)} 
\end{equation}

\noindent where all energies are normalized by 100~keV. The BATSE database  
includes the spectral data at 2048 ms time scale, which  
could be used to find the average spectral parameters at peak time intervals. For  
the group of bright BATSE bursts, the average spectral parameters at the peaks  
are $<\alpha>=-0.618$, $<E_{peak}>=329$~keV and $<\beta>=-3.18$  
(Mitrofanov {\it et al.} 1997).  
 
According to the concept of standard sources, one could use the average spectra  
of bright bursts $\phi^{(bright)}(E)$, as a standard distribution of photons for all  
emitters. Therefore, one might derive a universal relation between the observed  
photon fluxes $F$ and red-shifts $z$ of corresponding emitters. For two  basic
groups of 296 bright and 332 dim bursts with average peak fluxes  
$<F_{max}^{(bright)}>=6.15 \pm 0.35$ photons cm$^{-2}$ s$^{-1}$ and  
$<F_{max}^{(dim)}>=0.53 \pm 0.03$ photons cm$^{-2}$ s$^{-1}$,  
respectively,  this  
relation corresponds to the ratio

\begin{equation} 
\displaystyle 
\frac{ 
<F_{max}^{(bright)}> 
} 
{<F_{max}^{(dim)}> 
}= 
\frac{ 
\int^{E_{2}}_{E_{1}}\phi^{(bright)} [E(1+z_{br})]dE\cdot R^2(z_{dim}) 
} 
{ 
\int^{E_{2}}_{E_{1}}\phi^{(bright)} [E(1+z_{dim})]dE\cdot R^2(z_{br}) 
}      
\end{equation} 
 
\noindent where 
 
\begin{equation} 
R= 
\frac{c}{(1+z)q^{2}_{0}H_{0}}\cdot [q_{0}z+ 
(1-q_{0})(1-\sqrt{1+2zq_{0}})]      
\end{equation} 
 
\noindent is cosmological distance to a source, $H_{0}$ is the Hubble constant  
and $q_{0}$ represent the type of cosmological geometry. The geometry of the  
Universe with critical density is tested below with $q_{0}=\sigma_{0}=0.5$. The  
peak flux (photons cm$^{-2}$ s$^{-1}$) was calculated in the 50-300kev energy  
range according to 3B Catalog database. 
 
Using the average values $<F_{max}^{(bright)}>$ and $<F_{max}^{(dim)}>$  
and the average spectral law $\phi^{(bright)}(E)$, Exp. (16) could be  
transformed into the relationship between two cosmological parameters: an  
average red-shift $z_{dim}$ of emitters of dim bursts and an average stretching  
factor $Y$ between dim and bright  
bursts. Therefore, the $z_{dim}$ value could be implemented into the model  
function (11) $f_{dim}(t,E; z_{dim})$, as a free parameter, to check the  
consistency between the basic cosmological model and observed  
ACE$_{dim}^{(i)}$ curves for dim bursts.  
 
To do this, one has to put the $z_{dim}$ value into the model function  
$f_{dim}(t,E; z_{dim})$ and to calculate the difference (6) between the model and  
the ACE$_{dim}^{(2)}$ profile at the energy discriminator channel $i=2$. The estimated value  
$S_{dim}(z_{dim})$ could be corresponded to the probability $P_{300}(S_{(I-II)})$ (Figure 2) to find 
the difference $S_{(I-II)}$ equal to this value. The integrated  probability 
 
\begin{equation} 
P_{300}(z_{dim})=\int_{S_{dim}(z_{dim})}^{\propto}P_{300}(S_{(I-II)})dS_{(I-II)} 
\end{equation} 
 
\noindent could be interpreted, as the probability that cosmological model with $z_{dim}$ is consistent 
with observed  
ACE$_{dim}$ profile. Changing $z_{dim}$, one might create this way  
the probability function $P_{300}(z_{dim})$ (Figure 4).  
 
To take into account errors in the parameters of the basic analytical model of  
$f_{bright}(t,E)$, the main theoretical model (11)  
was used together with 12 additional models with ${\pm \sigma}$ deviations  
from the best fitting parameters (3). They compose the ${1\sigma \it corridor}$  
of models around the medium curve which corresponds to the best one (see  
Figure 4). 
 
It was found that the probability  $P(z_{dim})$ decreases with increasing  
$z_{dim}$ becoming as small as the level $\sim 3\cdot 10^{-3}$ of  
3${\sigma}$ fluctuations at $z_{dim}= 0.07-0.09$ (Figure 3). Therefore, one  
might consider the value $\sim 0.1$, as the 3${\sigma}$ upper limit for average  
red-shift of emitters of the basic group of 332 dim bursts with $F_{max}^{(1024)}<1$ photons  
cm$^{-2}$ s$^{-1}$. 
 
Two groups of bright and dim bursts are used for this estimation which been  
separated by the peak flux $F_{max}^{(1024)}=1$ photons cm$^{-2}$ s$^{- 
1}$. In this case one has the largest possible number of events in each sample,  
$\sim 300$, with the ratio of corresponding average peak fluxes of two samples   
$\sim 12$. However, one might suspect that selected sample of $\sim 300$ bright  
bursts might contain a large deal of bursts at cosmological distances, and, as  
such, a time dilation between bright and dim samples could be difficult to resolve.  
 
Formally speaking, this statement is not correct: in accordance with the basic  
cosmological model, the increase of  $z_{bright}$ value for the bright group  
results to more and more pronounced cosmological stretching of bursts from  
the dim sample, provided the ratio of their average peak fluxes is fixed. Indeed, the Exp. (16) points  
out that for a given ratio of peak fluxes  
$<F_{max}^{(bright)}>/<F_{max}^{(dim)}>$ an increase of $z_{bright}$ from  
the value 0 leads to increase of stretching factor $Y(z_{bright},  
z_{dim})$. Thus, for the  ratio  
$<F_{max}^{(bright)}>/<F_{max}^{(dim)}>=11.6$ one might find  
$Y=1.2, 1.6$ and 1.8 and $z_{dim}=0.3, 1.1$ and 1.7 for $z_{bright}=$0.1, 0.3 and 0.5, 
respectively.  
 
The found $3\sigma$ upper limit  $z_{dim} \sim 0.08$ corresponds to the  stretching  
factor $Y \sim 1.05$ and $z_{bright}=0.03$. One might conclude that the basic  
cosmological model with $\sim 300$ standard emitters of bright and dim bursts is  
consistent with the observed ACE$_{bright}$ and ACE$_{dim}$ curves,  
provided their red-shifts are $z_{bright}<0.03$ and $z_{dim}<0.1$, respectively. 
 
However, even taking into account the argument above, one could apply the  
proposed redshifting technique to perform a more conservative comparison  
between two samples of bright and dim bursts, which could be selected by a  
more stringent criterion based on the slope of $logN-logF$ distribution, and which  
would be truly isolated one from another by the sample of intermediate events in  
between.    
 
Let us select  two samples of 102 brightest bursts with    
$F_{max}^{(brightest)}>4.0$ photons cm$^{-2}$ s$^{-1}$ and 100 dimmest  
events with $F_{max}^{(dimmest)}< 0.41$ photons cm$^{-2}$ s$^{-1}$. The  
brightest sample corresponds to the -3/2 part of the $logN-logF$ distribution (see  
3B catalog, Meegan {\it et al.} 1995). There is about $\sim 400$  bursts with  
medium peak fluxes in between the brightest and the dimmest samples, and the  
ratio of the average peak fluxes   
$<F_{max}^{(brightest)}>/<F_{max}^{(dimmest)}>=49.8$ is as large as  
possible to imply the largest cosmological stretching  between them.  
 
For the new sample of the brightest 102 bursts the analytical approximation (2)  
corresponds to the best fitting parameters, which all agree quite well with the  
estimated $\pm 1\sigma$ corridor with the basic sample of 296 bright bursts.  
The best fitting parameters for the ACE$_{brightest}$ curve are  
$t_{brightest}^{(0)}= 1.88$ s,  
$\tilde{a}_{RF}^{(0)}=1.37$,  
$\tilde{a}_{BS}^{(0)}=1.26$, $\tilde{\alpha}_{1}= -0.25$,  
$\tilde{\alpha}_{2} = 0.065$ and  $\tilde{\alpha}_{3}=0.075$. Similarly to (11), these  
new parameters could be used to build the trial function $f_{dimmest}(t,E;  
Y(z_{brightest}, z_{dimmest}))$ (11) to fit the ACE$_{dimmest}$ curve of the sample  
of 100 dimmest bursts. 
 
The best fitting values of $Y^{**}$ equal  1.01, 0.80 and 0.88 for  
ACE$^{(i)}_{dimmest}$ at the three energy discriminators with numbers $i=$1,2 and  
3, respectively. The corresponding values of reduced $\chi^{2}$ are 2.95, 3.20 and 1.90,  
respectively. Therefore, the best fitting stretching factors $Y^{**}$ between the  
samples of the dimmest and the brightest bursts do not manifest any {\it stretching}. These values are similar 
to the best fitting factors between the basic 
samples of $\sim 300$ bright and dim bursts, and they all are consistent with the  
absence of any cosmological stretching.  
 
However, to find the upper limit of the stretching factors between the two samples of   
brightest and dimmest events, one has to compare the trial model  
$f_{dimmest}(t,E; z_{dimmest})$ (Exp. 11) with the ACE$_{dimmest}$ curve  
taking into account the sampling statistics of two groups. The probability  
distribution $P_{100}(S_{(I-II)})$ has to be used for the two sets of $\sim 100$  
events (see Section 3). According to Mitrofanov {\it at al.} (1997), the distribution  
of $P_{100}(S_{(I-II)})$ will have the same shape as the distribution $P_{300}(S_{(I-II)})$ for sets with 
$\sim 300$  
events. Therefore, the value of $S_{(I-II)}$ for the $3\sigma$ limit will be about  
the same. However, because of smaller statistics, for samples with $\sim  
100$ events the function (8) has denominator in $\sim 3$ times larger than for  
samples with $\sim 300$ events, and therefore, the difference between two ACEs  
profiles allowed by $3\sigma$ limit could be in $\sim 1.7$ times larger. 
 
Similarly to the basic case of two samples of $\sim 300$ bursts, the new samples  
of $\sim 100$ brightest and dimmest events are compared by the proposed  
technique, when for selected values of $z_{dimmest}$ the probability  
$P_{100}(z_{dimmest})$ is estimated (see (18)) to get the found difference  
between the model profile $f_{dimmest}(t,E; z_{dimmest})$ and  the observed  
ACE$_{dimmest}$ curve at the third energy discriminator channel. The  
corresponding probability function $P_{100}(z_{dimmest})$ is presented at the  
Figure 5. The $3\sigma$ upper limit of the $z_{dimmest}$ value is 0.46.   
 
Thus, when two samples of the brightest and the dimmest bursts with $\sim 100$  
events are compared, no significant increase is found for the best fitting stretching  
factors in comparison with the case of two basic samples of $\sim 300$ bright and  
dim bursts. At both cases one does not see any evidence for stretching effect at  
all. Using the sampling statistics of bursts, the $3\sigma$ upper limits are  
estimated of $z_{dim}$ for $\sim 300$ dim bursts and of $z_{dimmest}$ for  
$\sim 100$ dimmest events, which equal  $\sim 0.1$ and $\sim 0.5$,  
respectively. One could suspect that the larger upper limit in the second case   
results from the smaller sampling statistics of groups of $\sim 100$ bursts, and it  
hardly provides more evidence for cosmological stretching in comparison with the  
basic case of groups of $\sim 300$ events.  
 
However, formally speaking, one has to conclude that the basic cosmological  
models with standard candles are still allowed for gamma-ray bursts provided  
they correspond to the $3\sigma$ upper limit $z_{dim}<0.1$ for the group of dim  
bursts with $F_{max}^{(dim)}< 1.0$ photons cm$^{-2}$ s$^{-1}$ or to the  
upper limit $z_{dimmest}<0.5$ for the group of the dimmest  bursts with   
$F_{max}^{(dimmest)}< 0.41$ photons cm$^{-2}$ s$^{-1}$. These limits   
resulted from the different sampling statistics of these groups, and further  
observations of bursts will allow  either to decrease these limits, or to resolve  
the time-stretching effect of dim gamma-ray bursts with respect to bright ones.   
 
Two different  average photon spectra with power laws ${\alpha}=1$ and  
${\alpha}=2$  were used for the test of the basic samples also. At the plane  
$P_{300}(z_{dim})$  
versus $z_{dim}$ these models correspond to upper and lower lines around  
the main curve, which was found for the average energy spectra (Figure 6).  
Therefore, the shape of the photon energy spectra  does not affect significantly  
the upper limit of $z_{dim}$. The upper limits of $z_{dim}$ could be estimated  
also for different parameters of  cosmological geometry. Two curves for chance  
probability $P_{300}(z_{dim})$ were derived for two different sets of  
cosmological parameters (Figure 7): $q_{0}=\sigma_{0}=0.1$ (open Universe)  
and $q_{0}=\sigma_{0}=1.0$ (closed Universe). One might see that  these cases  
of the Universe geometry lead to 3${\sigma}$ upper limits    
$z_{dim}\sim0.08$ about the same as the case of flat expanding Universe  
($q_{0}=\sigma_{0}=0.1$) . 
 
\section{Discussion and Conclusions} 
 
So, the performed comparison of the ACE profiles for groups of bright and dim  
bursts  
does not allow $z_{dim}$  larger than $\sim0.1-0.5$ for the basic  
cosmological model with standard sources. Moreover, the ${\it ACE-based}$ 
limit of red-shift of dim bursts  agrees with non-cosmological models of GRBs in  
the flat Euclidean space. 
 
There are two well-known estimations of the red-shifts of emitters of GRBs  
according to cosmological models. The first one is based on the average  
parameter $<V/V_{max}>=0.33 \pm 0.01$ for 3B data base (Meegan {\it et al.}  
1995). One should expect to have $<V/V_{max}>=0.50$ for homogeneous  
distribution of standard sources in the Euclidean space. On the other hand,  
the observed parameter $<V/V_{max}>$ is consistent with the non-Euclidean  
geometry of expanding Universe. For distant emitters of dim bursts the ${\it  
geometry-based}$ upper limit of red-shift was estimated about 0.5-2.0   
(Wickramasinghe {\it et al} 1993). Taking into account the coupling between the  
spectral shape and the temporal profiles of bursts, Fenimore and Bloom (1995)  
have obtained much larger upper limit ${\sim}$2-6. 
 
Another estimation of cosmological limit of the red-shift was based on the effect  
of  
hardness/intensity correlation of GRBs. The average peak of $\nu F_{\nu}$  
spectra of dim bursts was found to be much softer than the average peak  
of bright bursts (Mallozzi {\it et al.} 1995). The corresponding ratio between  
peak energies of dim and bright bursts leads to the ${\it spectra-based}$ upper  
limit of red-shift, which was estimated about $\sim 1$. 
 
There is an agreement, at least qualitative, between ${\it geometry-based}$ and  
$\it spectra-based$ upper limits of red-shifts of distant emitters of GRBs. These  
estimations result to  $z_{dim}\sim 1$ or even much larger. On the other hand,  
the ${\it ACE-based}$ upper limit  of $z_{dim}\sim 0.1-0.5$  
does not agree with either the ${\it geometry-based}$ or the $\it spectra- 
based$ limits. Therefore, the basic model of GRBs with standard cosmological  
sources is not with all available constraints. This is the main conclusion of the present paper.

Developing a cosmological model of GRBs, one should postulate
some kind of $z$-dependent property(es) of outbursting sources which could
ensure the consistency. Generally speaking, $z$-dependence could be
attributed to different properties of  bursts sources, such as outbursts rate 
density, bursts luminosity, average energy spectra  and average light curves. 
There is a reasonable consistency between ${\it geometry-based}$ and ${\it 
spectra- based}$ limits of red-shifts of dim bursts emitters. Therefore, one 
might not suggest any intrinsic $z$-dependence either for the outburstsÆ rate 
density, or for the energy spectra of the emitted gamma-rays, because they
would both lead to consistent limits of $z$ for the model with
standard sources. On the other hand, to make the agreement
between them and the ${\it ACE-based}$ limit, one could postulate
some sort of intrinsic evolution of outbursting sources which
leads to intrinsic squeezing of their light curves with
increasing red-shifts. There are physical conditions in local
cosmological space which vary with $z$: the local density of
matter, the local temperature of microwave background, etc., but
at the present time no one knows how much these conditions could
actually influence on bursts' light curves, if they could at all.
Obviously, {\it a priori} there is no physical reason to propose
this kind of evolution, and it could be considered as a pure
phenomenological speculation.

In addition to ${\it ACE-based}$ test, the time-dilation tests
should also be done with another time-based parameters of bright
and dim GRBs, such as pulse width, interpulse duration, etc.
Comparison of distinct ${\it time-based}$  signatures for
different intensity groups of bursts would allow to distinguish
the basic effect of cosmological time-stretching and energy red-
shifting,  which should be identical for all time-energy
signatures,  from another effects resulted from $z$-dependent
evolution, which should be different for each of temporal
parameters. The cosmological paradigm of GRBs could be finally
approved at these tests, and new knowledge would be obtained about
intrinsic properties of close and distant GRBs sources in the co-
moving reference frames. This studies will be done elsewhere.

\section{Acknowledgments} 
 
We would like to thank members of BATSE team Drs. G.J. Fishman, W.S.  
Paciesas, M.S. Briggs, C.A.Meegan, R.D. Preece, G.N. Pendleton, J.J. Brainerd for fruitful cooperation which made possible this  
paper to be written.

This work was supported by the RFBR 96-02-18825 grant in Russia.
 
%
\clearpage 
 
\begin{deluxetable}{lccccc} 
\tablecaption{Best fitting factors $Y^{*}$ for ACE$_{dim}^{(i)}$ } 
 \tablehead{ 
\colhead{Energy range, keV}              & 
\colhead{Best fit}                  & 
\colhead{Reduced $\chi^2$}                    & 
\colhead{$P(>\chi^{2})$}            & 
\colhead{DOF}              & 
} 
\startdata 
25-50  & 0.81$\pm 0.02$ & 2.3 & $9.2\cdot 10^{-6}$ & 37 \nl 
50-100  & 0.86$\pm 0.01$ & 1.28 & 0.12 & 37 \nl 
100-300  & 0.82$\pm 0.02$ & 2.0 & $3.0\cdot 10^{-4}$ & 37 \nl 
25-300  & 0.84$\pm 0.02$ & 2.0 & $<10^{-6}$ & 113 \nl 
\enddata 
 
\end{deluxetable}

%

\clearpage 
\figcaption{At the left line three viewgraphs for ACE$_{bright}$ profiles  
are presented for three discriminator channels number $i=1,2,3$. The  
best fitting approximation profiles (2) are shown for each ACE. At the  
bottom viewgraph ($i=3$) the model for $i=1$ is shown ({\it dash line}) to  
demonstrate the energy dependence of ACE. At the right line three  
zooms of corresponding ACEs are presented to show the quality of  
approximations and the boundaries of ${\pm \sigma}$ {\it corridor} of  
models around the best ones.\label{Fig1}} 
\figcaption{The probability distribution of $S^{(2)}$ values divided by number of 
ACE bins (38) for energy discriminator $i=2$. It is provided by  
$10^{5}$ random choices of two groups of 303 bursts among the total  
set of 3B database. The  value corresponding to $3\sigma$ standard deviation 
is shown by {\it dotted-dashed} line. One sees that sampling statistics allows   
much larger difference between two sets of bursts than it could be  
expected from the sample variance for each of them.\label{Fig2}}  
 
\figcaption{ACE$^{(i)}_{dim}$ profiles for 332 dim bursts with  
$F^{(1024)}_{max}<1$ photons cm$^{-2}$ s$^{-1}$. The best fitting models   
(11) of ACEs at each discriminator channel are shown by solid lines. \label{Fig3}} 
\figcaption{The estimated probability $P_{300}(z_{dim})$ of consistency  
is shown between the model curve (11) based on the ACE$_{bright}$ profiles  
and the standard cosmological model and the  observed ACE$_{dim}$  
profile at the second ($i=2$) discriminator channel. The dashed region  
represents the probabilities for ${\pm \sigma}$ {\it corridor} of models  
around the best one. The {\it dotted-dashed} line shows the probability level  
of standard 3${\sigma}$ fluctuations.\label{Fig4}} 
\figcaption{The estimated probability $P_{100}(z_{dim})$ of consistency  
is shown between the model curve (11) based on the ACE$_{brightest}$ profiles  
and the standard cosmological model and the  observed ACE$_{dimmest}$  
profile at the third ($i=3$) discriminator channel. The {\it dotted-dashed} line shows the probability level  
of standard 3${\sigma}$ fluctuations.\label{Fig5}}

\figcaption{The same probability $P_{300}(z_{dim})$ is shown as at the Figure 4 (solid  
line) together with two another estimations based on the energy  
spectra with power law: {\it dashed} and {\it dotted} lines correspond to  
${\alpha=1}$ and ${\alpha=2}$, respectively.\label{Fig6}} 
\figcaption{The same probability $P_{300}(z_{dim})$ is shown as at the Figure 4 (solid  
line) together with two estimations corresponded to another  
models of Universe: {\it dashed} and {\it dotted} lines correspond to  
$q_{0}=0.1$ and $q_{0}=1.0$, respectively.\label{Fig7}}

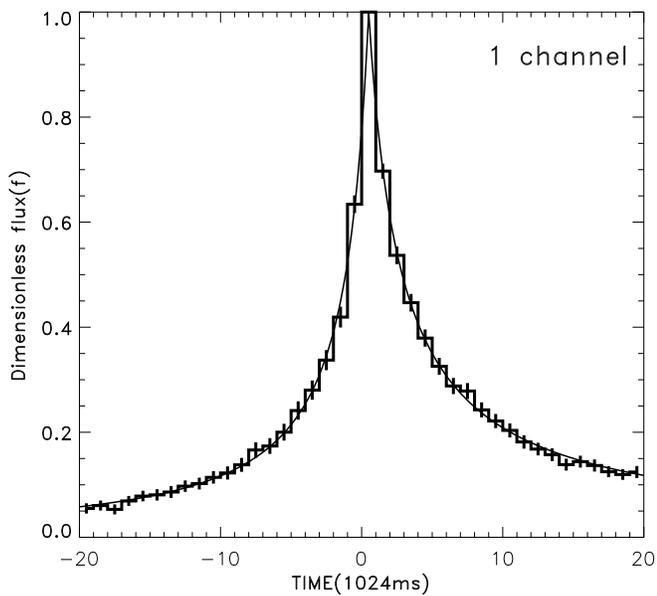
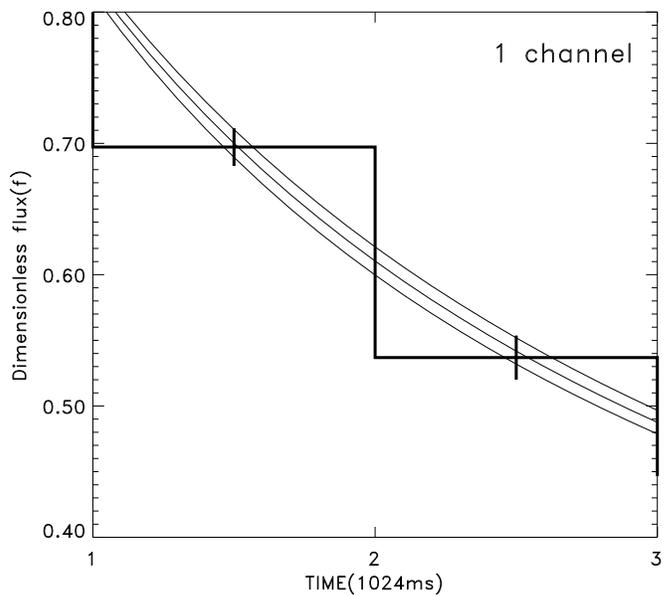
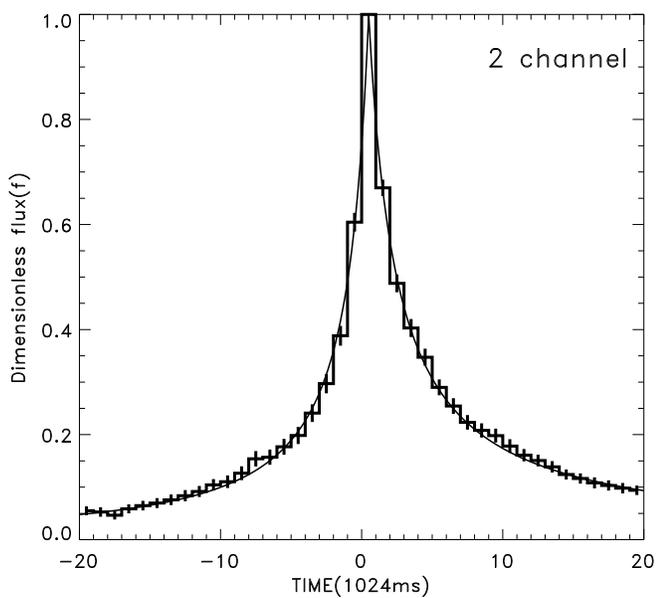
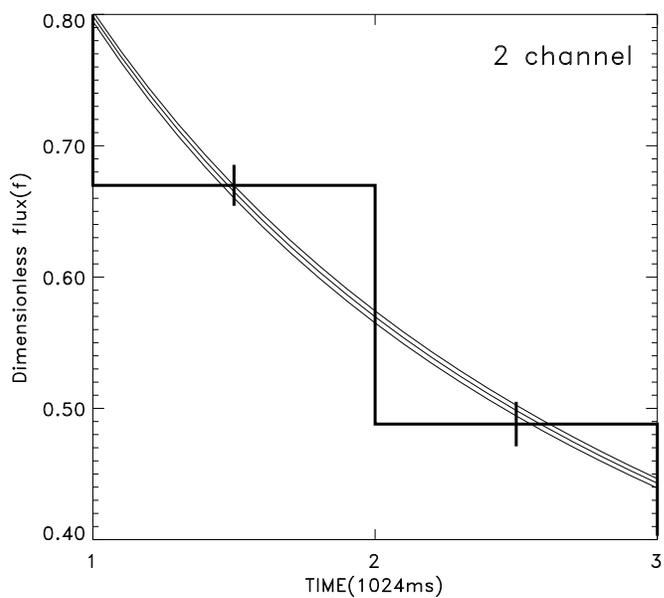
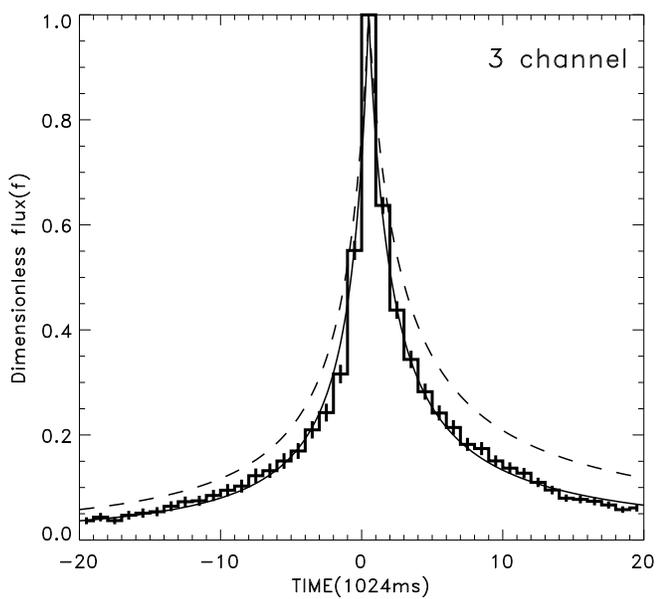
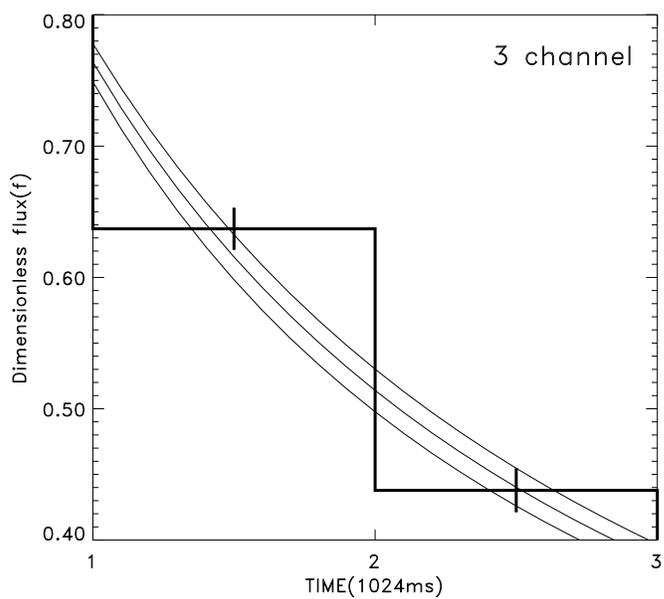
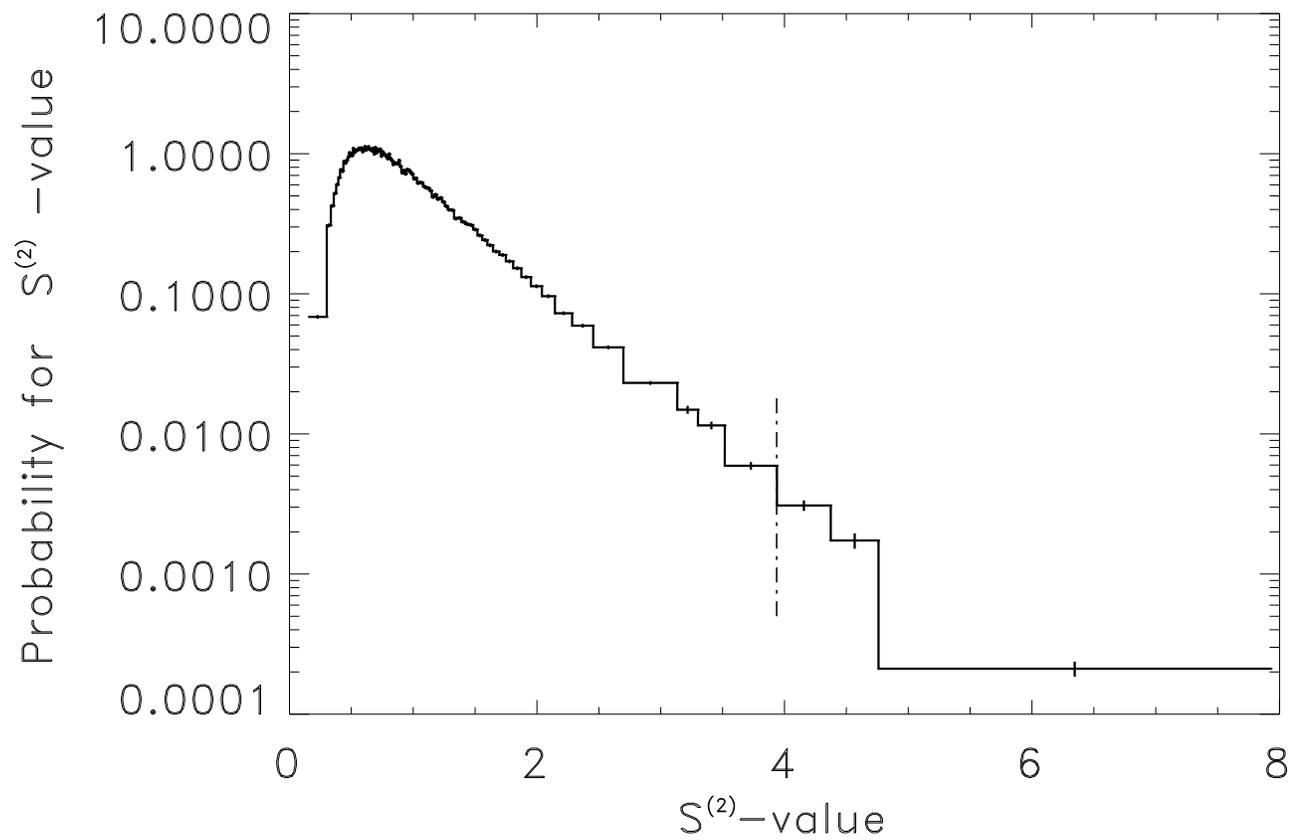
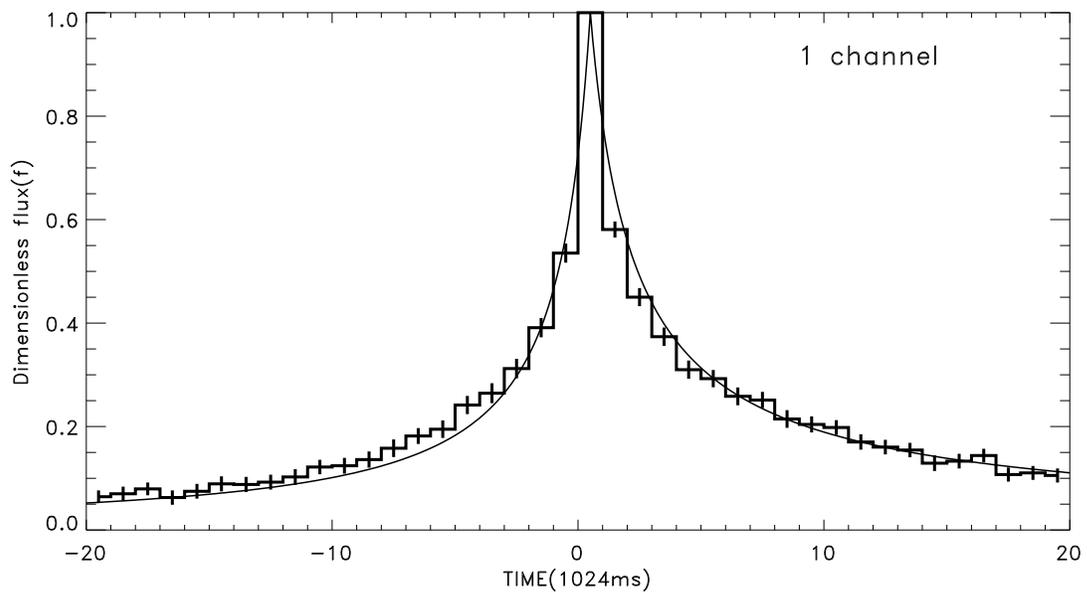
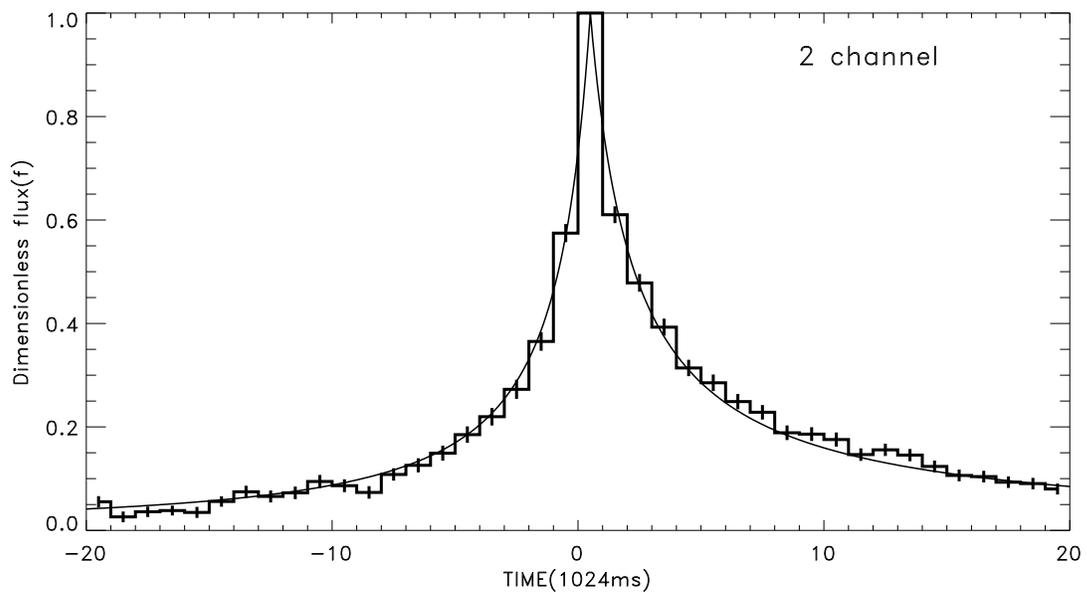
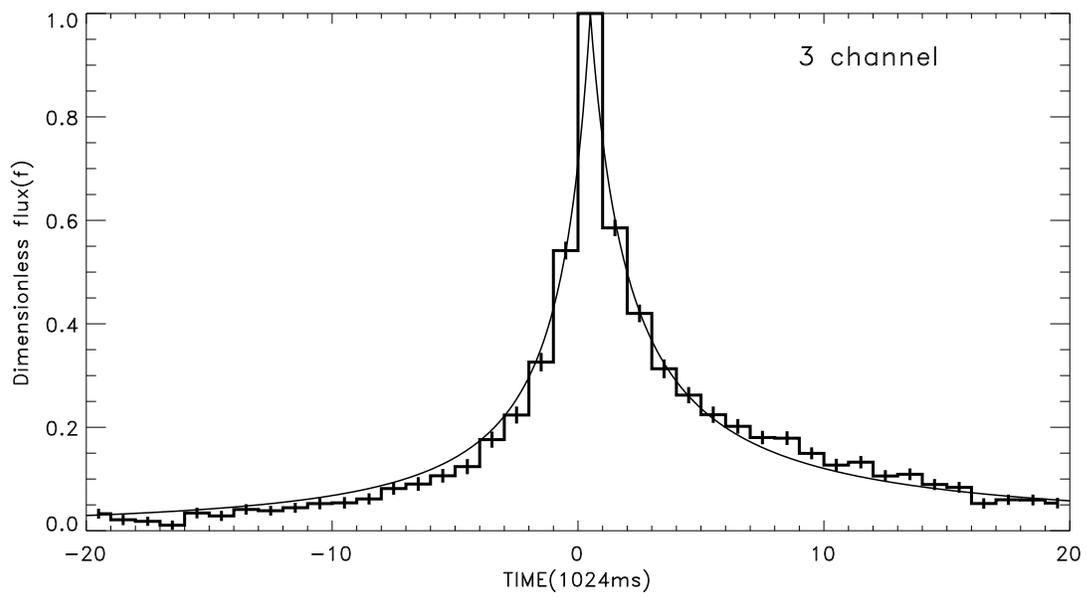
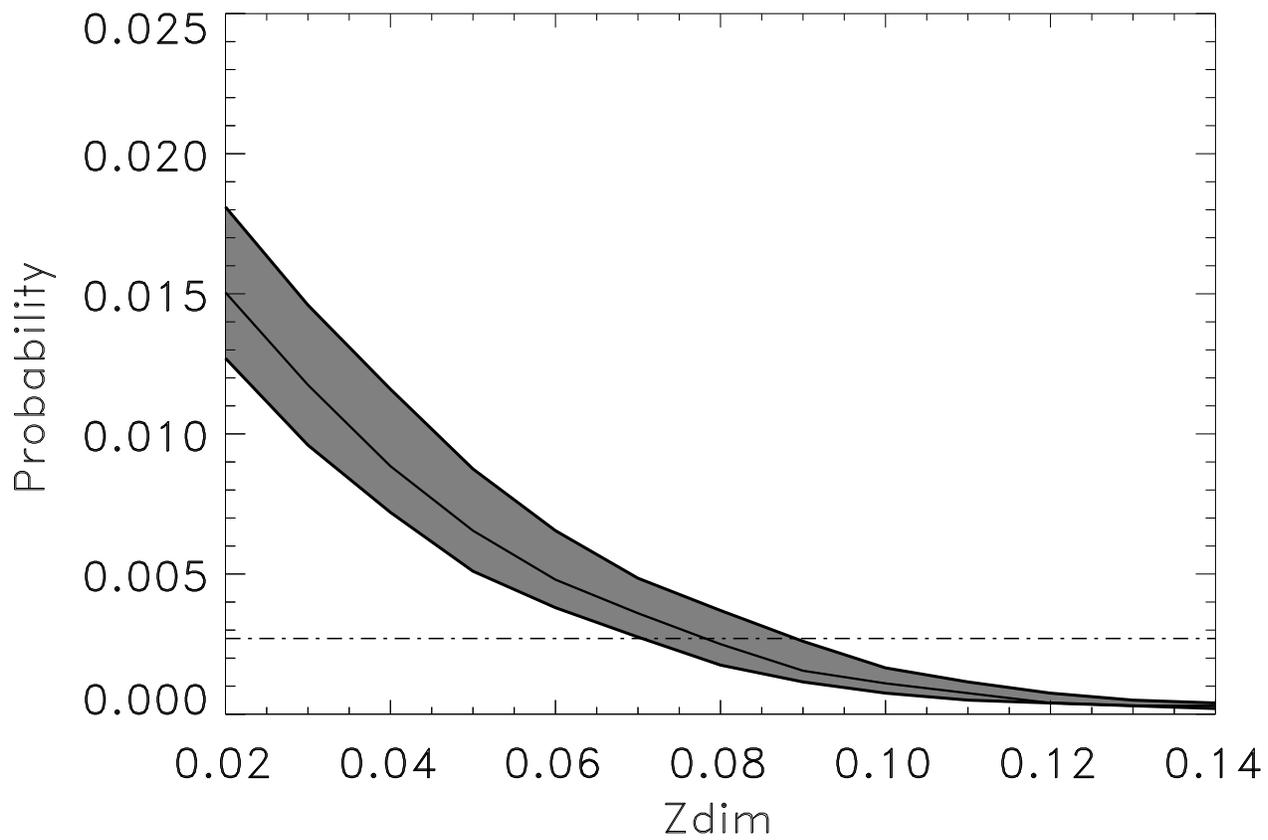
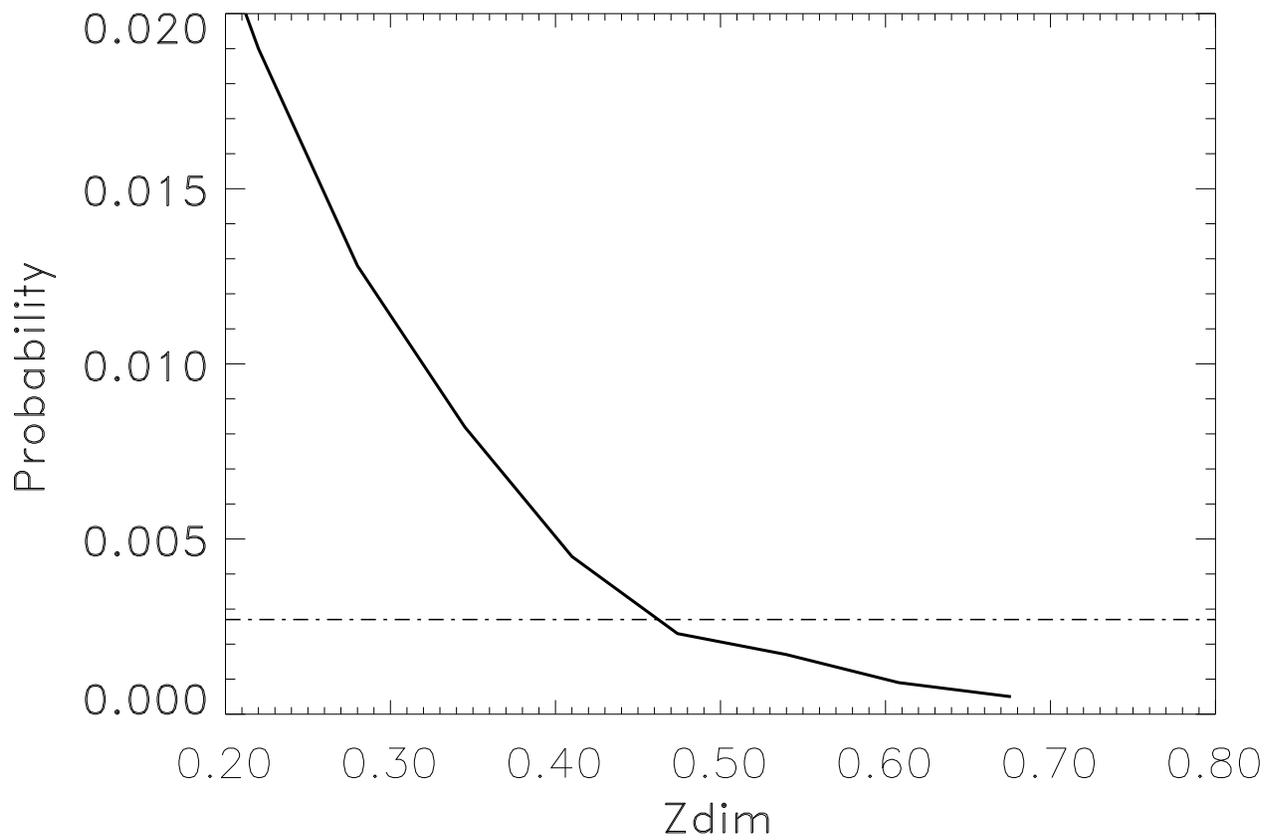
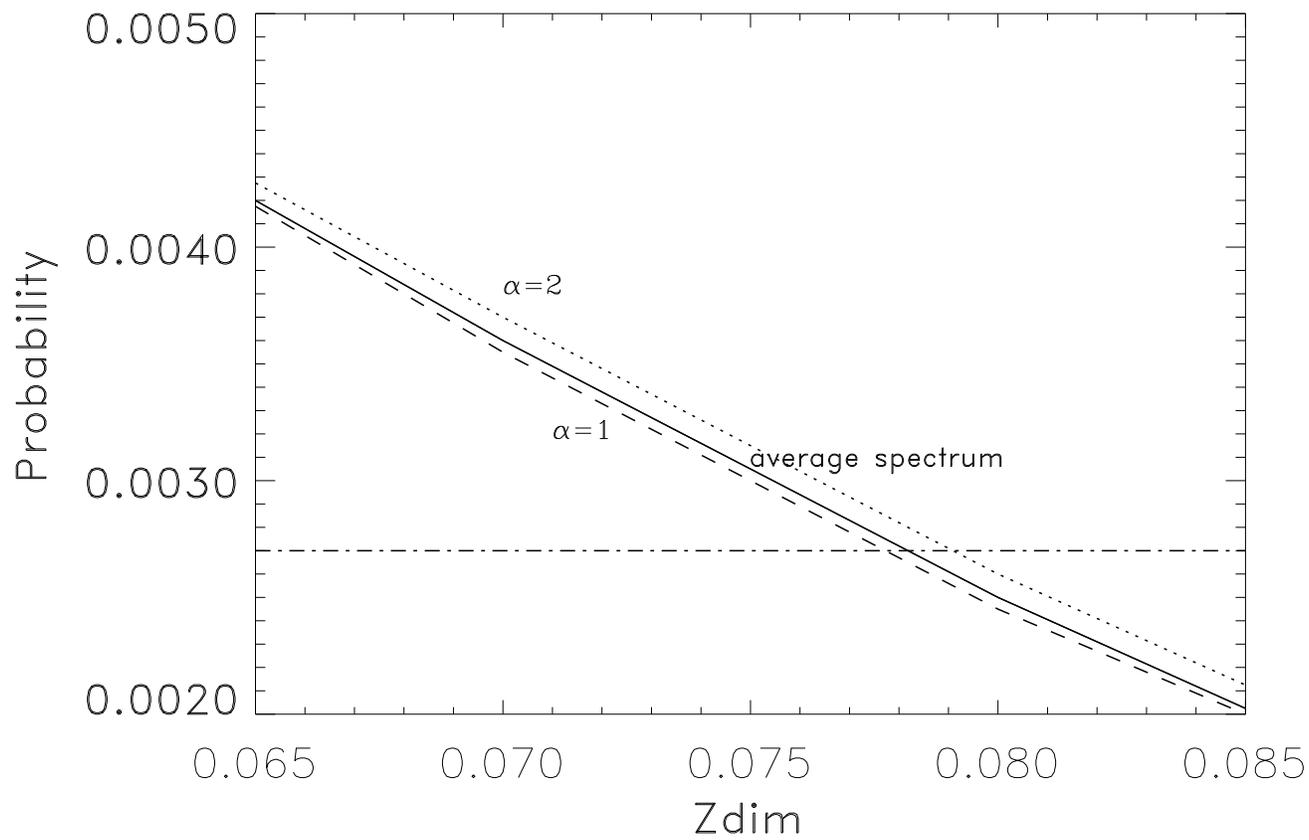
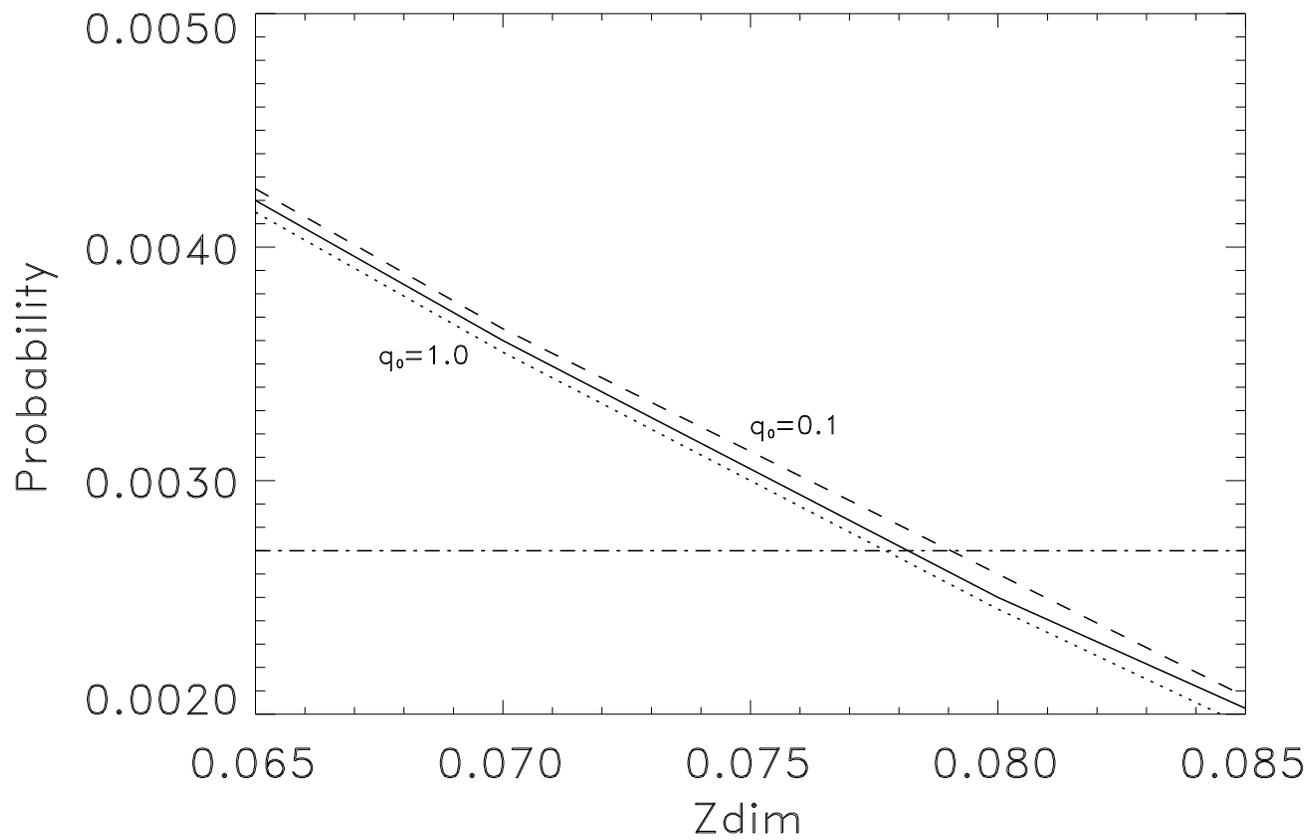
\end{document}